\documentclass[12pt,letterpaper]{article}  

\usepackage[includeheadfoot,
            marginratio={1:1,2:3}, 
            width=412pt, 
            height=688pt,]{geometry}

\usepackage{amsmath}
\usepackage{amsfonts}
\usepackage{amssymb}
\usepackage{stmaryrd}
\usepackage{empheq}
\usepackage[arrow,matrix,curve]{xy}

\newcommand{\nc}{\newcommand}

\nc{\lb}{\llbracket}
\nc{\rb}{\rrbracket}
\nc{\gl}{\llbracket}
\nc{\gr}{\rrbracket}
\newcommand{\eq}[1]{\begin{equation}
                     \begin{split} #1 \end{split}
                     \end{equation}}
\newcommand{\tri}{\hspace{-3.5pt}\vartriangle\hspace{-3.5pt}}
\newcommand{\ul}{\underline}

\begin{document}

\vspace*{-1.5cm}
\begin{flushright}
  {\small
  MPP-2012-86\\
  ITP-UU-12/16\\
  SPIN-12/14\\
  }
\end{flushright}

\vspace{2cm}
\begin{center}
{\LARGE
Bianchi Identities for Non-Geometric Fluxes \\
}
\vspace{0.4cm}

{\bf - From  Quasi-Poisson Structures to Courant Algebroids -
}
\end{center}

\vspace{0.35cm}
\begin{center}
  Ralph Blumenhagen$^{1}$, Andreas Deser$^{1}$, Erik Plauschinn$^{2}$ and 
Felix Rennecke$^{1}$
\end{center}

\vspace{0.1cm}
\begin{center} 
\emph{$^{1}$ Max-Planck-Institut f\"ur Physik (Werner-Heisenberg-Institut), \\ 
   F\"ohringer Ring 6,  80805 M\"unchen, Germany } \\[0.1cm] 
\vspace{0.25cm}
\emph{$^{2}$ Institute for Theoretical Physics and Spinoza Institute, \\
Utrecht University, 3508 TD  Utrecht, The Netherlands}  \\

\vspace{0.2cm}

 \vspace{0.5cm} 
\end{center} 

\vspace{1cm}

%%%%%%%%%%%%%%%%%%%%%%%%%%%%%%%%%%%%%%%%%%%%%%%
%%%%%%%%%%%%%%%%%%%%%%%%%%%%%%%%%%%%%%%%%%%%%%%
%%%%%%%%%%%%%%%%%%%%%%%%%%%%%%%%%%%%%%%%%%%%%%%
%%%%%%%%%%%%%%%%%%%%%%%%%%%%%%%%%%%%%%%%%%%%%%%
%%%%%%%%%%%%%%%%%%%%%%%%%%%%%%%%%%%%%%%%%%%%%%%
%%%%%%%%%%%%%%%%%%%%%%%%%%%%%%%%%%%%%%%%%%%%%%%
%%%%%%%%%%%%%%%%%%%%%%%%%%%%%%%%%%%%%%%%%%%%%%%
%%%%%%%%%%%%%%%%%%%%%%%%%%%%%%%%%%%%%%%%%%%%%%%

\begin{abstract}
Starting from a (non-associative) quasi-Poisson structure, 
the derivation of a Roytenberg-type algebra is presented. 
From the Jacobi identities of the latter, 
the most general form of  Bianchi identities for fluxes
$(H,f,Q,R)$ is then derived.
It is  also explained how this approach 
is related to the mathematical theory of quasi-Lie  and
Courant algebroids. 
\end{abstract}

\clearpage

%%%%%%%%%%%%%%%%%%%%%%%%%%%%%%%%%%%%%%%%%%%%%%%
%%%%%%%%%%%%%%%%%%%%%%%%%%%%%%%%%%%%%%%%%%%%%%%
%%%%%%%%%%%%%%%%%%%%%%%%%%%%%%%%%%%%%%%%%%%%%%%
%%%%%%%%%%%%%%%%%%%%%%%%%%%%%%%%%%%%%%%%%%%%%%%
%%%%%%%%%%%%%%%%%%%%%%%%%%%%%%%%%%%%%%%%%%%%%%%
%%%%%%%%%%%%%%%%%%%%%%%%%%%%%%%%%%%%%%%%%%%%%%%
%%%%%%%%%%%%%%%%%%%%%%%%%%%%%%%%%%%%%%%%%%%%%%%
%%%%%%%%%%%%%%%%%%%%%%%%%%%%%%%%%%%%%%%%%%%%%%%

\section{Introduction}
\label{sec:intro}

One of the most distinctive features of string theory  certainly is
T-duality. Applying this transformation to configurations which are already 
well understood has led to new insight about the theory 
and to the discovery of new structures  such as  D-branes 
and  mirror symmetry for  Calabi-Yau three-folds.
More recently, T-duality has been applied to closed-string backgrounds
with non-vanishing three-form flux, resulting in configurations
not known previously in the framework of supergravity.
More concretely, starting from a background with $H$-flux and 
performing a T-duality transformation 
along a single direction of isometry gives  a 
configuration with  geometric flux $f$ \cite{Dasgupta:1999ss}.
After further T-dualities, as illustrated in 
\cite{Kachru:2002sk,Shelton:2005cf},  backgrounds with 
so-called non-geometric fluxes $Q$ and $R$ are obtained. This chain of 
transformations can be summarized by
\eq{
  H_{abc} \;\xleftrightarrow{\;\; T_{c}\;\;}\;
   f_{ab}{}^{c} \;\xleftrightarrow{\;\; T_{b}\;\;}\;
  Q_{a}{}^{bc} \;\xleftrightarrow{\;\; T_{a}\;\;}\;
  R^{abc} \; ,
}
where we would like to note that lower indices are form indices and upper ones
are vector indices.

For the case of $Q$-flux, the underlying structure can be understood using the
notion of T-folds \cite{Dabholkar:2002sy,Hull:2004in,Hull:2006va}. 
However, the nature of backgrounds with $R$-flux is less clear, as the manifold is
expected to  not even be locally geometric. In fact, in
\cite{Bouwknegt:2004ap,Bouwknegt:2004tr} it has been argued 
that these configurations have a non-associative structure.
Further evidence for this observation was obtained by pursuing a conformal field
theory (CFT) analysis \cite{Blumenhagen:2010hj,Lust:2010iy,Blumenhagen:2011ph}, 
where T-duality is  realized
via a reflection of right-moving coordinates. The main results of this work 
can be summarized by the following commutator and Jacobiator of the coordinates 
(see also \cite{Andriot:2012an})
\eq{
\label{cftnca}
            [x^i,x^j]= \oint_{C_x} Q_k{}^{ij} dy^k\;, \hspace{50pt}
           [x^i,x^j,x^k]= R^{ijk}\; .
}
These expressions imply that, depending on the fluxes $Q$ and $R$, the coordinates can 
be both non-commutative and non-associative (NCA). 
As can be seen from the first expression, 
the commutator of two coordinates is related to a Wilson line
and so only a string with a non-vanishing winding number can detect 
such a non-commutative structure.
This was shown in the recent work  \cite{Condeescu:2012sp}. 
On the other hand, as illustrated in \cite{Blumenhagen:2010hj}, 
a non-associative structure indicated by a non-vanishing Jacobiator does not depend on the winding 
number of the closed string.

Recently, double-field theory (DFT) \cite{Hull:2009mi} has been established
as  a manifestly T-duality invariant
formulation of the low-energy effective action of string theory.
As such, it provides a powerful
tool to analyze  non-geometric  fluxes (at least at tree-level).
The new ingredient in DFT is a bi-vector field $\beta$
which can be seen as the ``T-dual'' of the Kalb-Ramond field $B$, and
the fluxes $Q$ and $R$ are expressed in terms of $\beta^{ij}$ as
 $Q_k{}^{ij}=\partial_k \beta^{ij}$ and
$R^{ijk}=\beta^{im}\partial_m \beta^{jk}+{\rm cycl.}$\footnote{A bi-vector field
also appears in \cite{Halmagyi:2009te} and \cite{Andriot:2011uh}, where the expressions for the fluxes $Q$ and $R$ can be found as well.}. 
In \cite{Andriot:2012wx,Andriot:2012an} (see also \cite{Aldazabal:2011nj})
the double-field theory action for a non-vanishing
bi-vector field has been worked out explicitly. Furthermore,
in \cite{Andriot:2012wx} a Bianchi identity for the $R$-flux was found, which
was discovered independently  in \cite{Blumenhagen:2012ma} following a different
approach based on the Schouten--Nijenhuis bracket.
In the latter paper, two more Bianchi identities for the
non-geometric fluxes were presented.

\medskip
In this letter, our  central objection is to generalize  
the above-mentioned Bianchi identities to the most general case in which all
fluxes $(H,f,Q,R)$ are non-vanishing and non-constant. 
For constant fluxes a  derivation was given in \cite{Shelton:2005cf,Ihl:2007ah,Aldazabal:2008zza},
but here we aim for the derivative corrections to these  identities.
Supporting space-time dependent structure "constants", the theory
of Lie and Courant algebroids
has been been proposed in the literature (see e.g. \cite{Halmagyi:2008dr,Halmagyi:2009te,Hull:2009zb})
as a suitable framework for describing these fluxes.
After providing the relevant definitions, we show 
how the fluxes $(H,f,Q,R)$ fit into this scheme. In fact, they
allow to define two quasi-Lie algebroids, which combine
into a Courant algebroid. To verify the axioms
of the latter, the generalized Bianchi identities turn out
to  play an essential role.

%%%%%%%%%%%%%%%%%%%%%%%%%%%%%%%%%%%%%%%%%%%%%%%
%%%%%%%%%%%%%%%%%%%%%%%%%%%%%%%%%%%%%%%%%%%%%%%
%%%%%%%%%%%%%%%%%%%%%%%%%%%%%%%%%%%%%%%%%%%%%%%
%%%%%%%%%%%%%%%%%%%%%%%%%%%%%%%%%%%%%%%%%%%%%%%
%%%%%%%%%%%%%%%%%%%%%%%%%%%%%%%%%%%%%%%%%%%%%%%
%%%%%%%%%%%%%%%%%%%%%%%%%%%%%%%%%%%%%%%%%%%%%%%
%%%%%%%%%%%%%%%%%%%%%%%%%%%%%%%%%%%%%%%%%%%%%%%
%%%%%%%%%%%%%%%%%%%%%%%%%%%%%%%%%%%%%%%%%%%%%%%

\section{Non-geometric fluxes and quasi-Poisson structures}

As it is well-known, at the massless level the gravity sector of string theory 
contains the metric $g_{ij}$, the anti-symmetric Kalb-Ramond  
field $B_{ij}$ and the dilaton $\phi$. 
Starting then from a
geometric background with non-vanishing three-form flux $H=dB$
and applying T-dualities,  one  obtains
backgrounds which are no longer geometric.
For these non-geometric configurations,  
the degrees of freedom are  more conveniently described by 
a dual metric $\tilde g_{ij}$, an anti-symmetric  
bi-vector field $\beta^{ij}$ and a dilaton $\tilde\phi$. 
Defining  ${\cal E}_{ij}=g_{ij}+B_{ij}$
and  $\tilde{\cal E}_{ij}=\tilde g^{ij}+\beta^{ij}$, 
the relation between the two sets of fields is given
by $\tilde{\cal E}={\cal E}^{-1}$, which in  components reads 
\cite{Grana:2008yw,Andriot:2011uh}
(see also  \cite{Halmagyi:2008dr, Halmagyi:2009te}
for a derivation from  a  world-sheet
point of view)
\eq{
  \label{relation_dft}
           \tilde g_{ij} =g_{ij}- B_{im} g^{mn} B_{nj}\;, \hspace{50pt}
           \beta^{ij}=-g^{im} B_{mn} \tilde g^{nj} \;,
} 
and $\tilde\phi$ is defined via 
$\sqrt{-g} \exp(-2\phi)=\sqrt{-\tilde g} \exp(-2\tilde\phi)$.
In double-field theory, which is an explicitly $O(D,D)$-invariant framework
consisting of a space with doubled coordinates $(x^i, \tilde x^i)$, the relation
\eqref{relation_dft} is just a particular $T$-duality transformation 
\cite{Hohm:2010jy,Andriot:2012an}.

We therefore see that an anti-symmetric bi-vector field $\beta\in \Gamma(\Lambda^2TM)$  plays an important role for non-geometric fluxes. In coordinates, it can be written as
$\beta=\frac{1}{2} \beta^{ij}\, e_i\wedge e_j$, where  $e_i= \partial_i$ denotes a basis vector in $TM$.
Such a bi-vector induces  two new
structures: a quasi-Poisson structure and an anchor map, which we will discuss in the present 
section.

%%%%%%%%%%%%%%%%%%%%%%%%%%%%%%%%%%%%%%%%%%%%%%%
%%%%%%%%%%%%%%%%%%%%%%%%%%%%%%%%%%%%%%%%%%%%%%%

\subsubsection*{Quasi-Poisson structure and NCA geometry}

Given an anti-symmetric bi-vector field $\beta$,  one can
define  a \emph{quasi-Poisson} structure as follows
\eq{
\label{quasipoissona}
                \{ f, g\}=\beta^{ij}\, (\partial_i f)\, (\partial_j g) \;,
}
where $f,g \in \mathcal{C}^{\infty}(M)$.
In general, this bracket does not satisfy the Jacobi identity but one  finds 
\eq{
\label{quasipoissonb}
      \{\{ f, g\},h\}+{\rm cycl.}= R^{ijk}\,   (\partial_i f)\, (\partial_j g)\,
         (\partial_k h) \;.
}
Here, 
$R^{ijk}$ is given by\footnote{Here and in the following, underlined indices are anti-symmetrized and  anti-symmetrization is defined as
$A_{[\ul a_1 \ul a_2 \ul a_3 \ldots \ul a_n]} = {\textstyle \frac{1}{n!} } 
  \sum_{\sigma\in S_n} {\rm sign}(\sigma)\,
  A_{a_{\sigma (1)}\,  a_{\sigma (2)}\ldots a_{\sigma (n)}}$.
}
$R^{ijk}=3\,\beta^{[\ul{i} m}\,\partial_m \beta^{\ul{jk}]}$
which takes the form of the   non-geometric $R$-flux mentioned earlier.
For vanishing $R\in \Gamma(\Lambda^3TM)$, 
one obtains a Poisson structure.

Next, we consider the second non-geometric flux $Q$ 
which can be expressed as $Q_k{}^{ij}=\partial_k \beta^{ij}$, and which in general is not to be considered  a tensor.
To make contact with the CFT results \eqref{cftnca},
we write the quasi-Poisson structure  \eqref{quasipoissona}  for two 
coordinates $x^i$ and $x^j$ as
\eq{
\label{quasipoissonc}
                \{ x^i, x^j\}=\oint_{C_x} Q_k{}^{ij} dy^k\; .
}
That means, a non-trivial 
Wilson line of the $Q$-flux can 
detect a non-trivial Poisson bracket between two
coordinates.

Let us mention that both \eqref{quasipoissonb} and \eqref{quasipoissonc} can be considered
as the classical limits of the quantum non-commutativity and
non-associativity investigated   
from a conformal
field theory point of view in \cite{Blumenhagen:2010hj,Lust:2010iy,Blumenhagen:2011ph,Condeescu:2012sp}. For instance, in \cite{Blumenhagen:2011ph} it was found
that a string moving in a  background with constant $R$-flux gives
rise to a non-trivial three-product of the form
\eq{
  \label{triprod}
      f\,\tri\, g\,\tri\, h(x)=f\, g\, h +  R^{ijk} (\partial_i f)\,
            (\partial_j g)\, (\partial_k h) + \mathcal O(R^2)\; .
}  
The quantum version of \eqref{quasipoissonc} was considered
in \cite{Lust:2010iy,Condeescu:2012sp}.      
Therefore, in analogy to the Moyal-Weyl product, \eqref{triprod}
 points towards  the existence of a deformation quantization
of the classical quasi-Poisson structure. However, this certainly very interesting
question is beyond the scope of this letter.

%%%%%%%%%%%%%%%%%%%%%%%%%%%%%%%%%%%%%%%%%%%%%%%
%%%%%%%%%%%%%%%%%%%%%%%%%%%%%%%%%%%%%%%%%%%%%%%

\subsubsection*{Anchor map}

In addition to the quasi-Poisson structure, the bi-vector $\beta$ also
induces a natural map $\beta^\sharp: T^*M\to TM$ 
from the co-tangent to the tangent space of a manifold $M$. It is defined by the relation\footnote{This map is closely related to the $\beta$-transform in \cite{Grana:2008yw}.}
\eq{
\label{anchorm}
\beta^\sharp(\eta)(\xi)=\beta(\eta,\xi) 
\hspace{50pt}{\rm for~all~}\ \xi\in T^*M\, ,
}
and, as we will discuss later, such a map is called an anchor.
In components,  equation \eqref{anchorm} reads as follows: 
denoting by $\{ e^i\}\in T^*M$ the basis dual to $\{e_i\}$, that is $e^i(e_j)=\delta^i_j$,
one finds
\eq{
\label{anchorm_comp}
  e^i_\sharp=\beta^\sharp(e^i)=\beta^{ij}\partial_j \;.
}  
Thus, there are two kinds of  derivative operators:
$e_i$ and $e^i_\sharp$.
We also observe that  the derivative 
 \raisebox{0mm}[3mm]{$\tilde D^i=\tilde\partial^i - \beta^{ij}\partial_j$} 
introduced in \cite{Andriot:2012wx},  
in the case of vanishing winding derivatives $\tilde\partial^i$,
is related to \eqref{anchorm_comp} as
$\tilde D^i= - e^i_{\sharp}$.

It is now straightforward to show that the differential operators
$\{e_i,e_\sharp^i\}\in TM$  satisfy the following commutation relations (see \cite{Grana:2008yw} for closely related expressions)
\eq{
\label{liealgebraa}
       [e_i,e_j]&=0 \;, \\
       [e_i,e^j_\sharp]&=Q_i{}^{jk}\, e_k \;, \\
       [e_\sharp^i,e_\sharp^j]&=R^{ijk}\, e_k + Q_k{}^{ij}\, e^k_\sharp \;,
}
where
$R^{ijk}=3\,\beta^{[\ul{i} m}\,\partial_m \beta^{\ul{jk}]}$
and 
$Q_k{}^{ij}=\partial_k \beta^{ij}$ as defined above.
The Jacobi identities for this Lie bracket imply  Bianchi identities
for the non-geometric $Q$- and $R$-fluxes. For vanishing $H$-flux and vanishing
geometric flux $f$, they read
\eq{
\label{bianchia}
   0&=  3\hspace{1pt} \beta^{[\ul a m} \partial_m Q_d{}^{\underline{bc}]}
     -\partial_d R^{abc}
 + 3\hspace{1pt}  Q_d{}^{[\ul a m} Q_m{}^{\ul b \ul c]} \;, \\
   0&=  2 \hspace{1pt} \beta^{[\underline{a} m}\, \partial_{m}  {R}^{\underline{bcd}]}- 3\hspace{1pt}
          {R}^{[\underline{ab}m}\, {Q}_m{}^{\underline{cd}]}  \;.
}
These relations were  derived  in \cite{Blumenhagen:2012ma} using 
a  different approach based on the Schouten--Nijenhuis bracket.\footnote{Note that in 
\cite{Blumenhagen:2012ma} we used a convention for $R$ which differs by a factor of two.} 
The second identity also
appeared in \cite{Andriot:2012wx} in the context of double-field theory. 
In the next section, we  generalize the algebra 
\eqref{liealgebraa} and 
the Bianchi identities \eqref{bianchia} to the situation when all
fluxes $(H,f,Q,R)$ are non-vanishing.

%%%%%%%%%%%%%%%%%%%%%%%%%%%%%%%%%%%%%%%%%%%%%%%
%%%%%%%%%%%%%%%%%%%%%%%%%%%%%%%%%%%%%%%%%%%%%%%
%%%%%%%%%%%%%%%%%%%%%%%%%%%%%%%%%%%%%%%%%%%%%%%
%%%%%%%%%%%%%%%%%%%%%%%%%%%%%%%%%%%%%%%%%%%%%%%
%%%%%%%%%%%%%%%%%%%%%%%%%%%%%%%%%%%%%%%%%%%%%%%
%%%%%%%%%%%%%%%%%%%%%%%%%%%%%%%%%%%%%%%%%%%%%%%

\section{Bianchi identities for all fluxes }
\label{sec:Bianchi}

In order to implement the  geometric flux 
$f$ into our discussion, we have to introduce  
a vielbein basis. 
The notation we employ is the following.
The vielbeins are denoted by $\{e_a{}^i\}$, and we define in the usual way
\eq{
  \label{def_01}
  e_a = e_a{}^i \partial_i \;.
}
The $\{e_a{}^i\}$ are required to be orthonormal with respect to the metric $g_{ij}$ on $M$, that is $e_a{}^i \hspace{1pt} e_b{}^j \hspace{1pt} g_{ij} = \delta^{ab}$, and the inverse of $e_a{}^i$ is 
denoted by $e^a{}_i$.
Furthermore,  in general the fields $e_a$ do not commute but have a non-vanishing Lie bracket
\eq{
  \label{comm_01}
  [ e_a, e_b ] = f_{ab}{}^c \,e_c \;,
}
where $f_{ab}{}^c$ is called  geometric flux  and is given by  
$f_{ab}{}^c= e^c{}_j\,(e_a{}^i\hspace{1pt} \partial_i
e_b{}^j-e_b{}^i\hspace{1pt}\partial_i e_a{}^j)$.
The basis dual to $\{ e_a\}$ will be denoted by $\{ e^a\}$ and is constructed as
$e^a = e^a{}_i e^i$.
The map \eqref{anchorm} induced by the quasi-Poisson structure $\beta$ acting on $e^a$ then reads
\eq{
  \label{def_02}
  e_\sharp^a = \beta^\sharp (e^a)= e^a{}_i \,\beta^{ij} \,\partial_j = \beta^{ab} \hspace{1pt}e_b \;,
}
where we employed $\beta^{ab} = e^a{}_i e^b{}_j \beta^{ij}$. 
Recall  that for the case of a vanishing torsion tensor  we have the relation
\eq{    
  de^a= -\tfrac{1}{2}\, f_{bc}{}^a\, e^b\wedge e^c \, ,
}
and  the connection coefficients $\Gamma^c{}_{ab}$ in the
non-coordinate basis satisfy $f_{ab}{}^c = \Gamma^c{}_{ab} -
\Gamma^c{}_{ba}$. The corresponding covariant derivative is  denoted by
$\nabla_a$.

%%%%%%%%%%%%%%%%%%%%%%%%%%%%%%%%%%%%%%%%%%%%%%%
%%%%%%%%%%%%%%%%%%%%%%%%%%%%%%%%%%%%%%%%%%%%%%%

\subsubsection*{Pre-Roytenberg algebra}

After having introduced out notation for the vielbeins, let us now evaluate two Lie brackets. In particular, we compute
\eq{
  \label{comm_02}
  [ e_a, e_\sharp^b ] = Q_a{}^{bc} e_c - f_{ac}{}^b e_\sharp^c \;, \hspace{40pt}
  [e_\sharp^a, e_\sharp^b] = R^{abc} e_c + Q_c{}^{ab} e_\sharp^c \;.
}
The terms appearing on the right-hand side are given by 
\eq{
\label{QRdef}
  Q_a{}^{bc} &=\partial_a \beta^{bc}+ f_{am}{}^b\, \beta^{mc} -f_{am}{}^c\,
  \beta^{mb} \;, \\[1mm]
 R^{abc}&= 3\, \bigl( \beta^{[\ul am}\,\partial_m \beta^{\ul b\ul c]}+f_{mn}{}^{[\ul a}\, 
 \beta^{\ul bm}\beta^{\ul c]n} \bigr)
 = 3\, \beta^{[\ul am}\nabla_m \beta^{\ul b\ul c]} \;,
}
and  correspond to the non-geometric $Q$- and
$R$-fluxes in the presence of  non-vanishing geometric flux $f$. 
However, the remaining  flux $H\in \Gamma(\Lambda^3T^*M)$ does not yet appear in the 
Lie algebra defined by the commutators \eqref{comm_01} and \eqref{comm_02}.
Let us therefore  perform the following redefinitions 
\eq{
  \label{redef_01}
           {\cal H}_{abc}&=H_{abc} \;,\\
           {\cal F}_{ab}{}^c&=f_{ab}{}^c - H_{abm}\, \beta^{mc} \;, \\
           {\cal Q}_{a}{}^{bc}&=Q_{a}{}^{bc} +  H_{amn}\, \beta^{mb}\,
           \beta^{nc} \;,\\
           {\cal R}^{abc}&=R^{abc} - H_{mnp}\, \beta^{ma}\,
           \beta^{nb}\, \beta^{pc} \;,
}
where $H_{abc}$ denotes the components of the usual $H$-field in the non-coordinate basis. Recalling its definition $H= dB$ in terms of the two-form gauge field $B$, we can infer
\eq{
  \mathcal H_{abc}  = H_{abc} 
  = 3 \,\nabla_{[\ul a} B_{\ul b \ul c]}
  \;.
}  
Employing \eqref{redef_01}, the commutators \eqref{comm_01} and
\eqref{comm_02} then take the form (see also \cite{Grana:2008yw})
\eq{
\label{preroytenberg}
       [e_a,e_b]&= {\cal F}_{ab}{}^c\, e_c +  {\cal H}_{abc}\, e_\sharp^c \;, \\
       [e_a,e_\sharp^b]&= {\cal Q}_{a}{}^{bc}\, e_c -  {\cal F}_{ac}{}^b\, e_\sharp^c\;, \\
       [e_\sharp^a,e_\sharp^b]&= {\cal R}^{abc}\, e_c +  {\cal Q}_{c}{}^{ab}\, e_\sharp^c \;.
}
These relations are very similar to the Roytenberg bracket
\cite{Roytenberg:01,Roytenberg:02} for a particular basis 
(see \cite{Halmagyi:2009te}), with the only difference that
the latter is not given by a Lie bracket on $TM$ but rather
by a so-called Courant bracket on $TM\oplus T^*M$.
Therefore, we call  \eqref{preroytenberg}  a \emph{pre-Roytenberg algebra}, and we will clarify the relation between these two structures in the next section.

%%%%%%%%%%%%%%%%%%%%%%%%%%%%%%%%%%%%%%%%%%%%%%%
%%%%%%%%%%%%%%%%%%%%%%%%%%%%%%%%%%%%%%%%%%%%%%%

\subsubsection*{Bianchi identities}

Using the commutation relations \eqref{preroytenberg}, we can now 
deduce the  Bianchi identities
for the various fluxes from the Jacobi identities of the
pre-Roytenberg algebra. 
First, we have the usual Bianchi identity for the $H$-flux
which in the presence of geometric flux can be written as
\eq{
  \label{bianchi0}
   {\rm I}\ :\ 0=\nabla_{[\underline{a}} \, {\cal H}_{\underline{bcd}]}=
    \partial_{[\underline{a}} \, {\cal H}_{\underline{bcd}]} - \tfrac{3}{ 2}\,
    {\cal F}_{[\ul{ab}}{}^m\, {\cal H}_{m\ul{cd}]} \;.
}
Next, since the Lie bracket satisfies the Jacobi identity, we can evaluate the following four equations
\eq{
  \begin{array}{r@{\hspace{5pt}}clr@{\hspace{5pt}}cl}
  {\rm II} &: & 0 = \bigl[ [e_a, e_b ], e_c \bigr] + {\rm cycl.} \;, \hspace{40pt} &
  {\rm III} &:   & 0 = \bigl[ [e_a, e_b ], e_\sharp^c \bigr] + {\rm cycl.} \;, \\[2mm]
  {\rm IV}&:    &0 = \bigl[ [e_a, e_\sharp^b ], e_\sharp^c \bigr] + {\rm cycl.} \;, &
  {\rm V}&:    &0 = \bigl[ [e_\sharp^a, e_\sharp^b ], e_\sharp^c \bigr] + {\rm cycl.} \;,
  \end{array}    
}
which lead to the four Bianchi identities
\eq{
\label{bianchi}
 {\rm II}:\hspace{5pt} 0 = &   \Bigl( \partial_{[\underline{c}} \, {\cal F}_{\underline{ab}]}{}^d +
    {\cal F}_{[\underline{ab}}{}^m\, {\cal F}_{\underline{c}]m}{}^d +
    {\cal H}_{[\underline{ab}\,m}\, {\cal Q}_{\underline{c}]}{}^{md}\Bigr)\\
   & \hspace{15pt}+\Bigl( \partial_{[\underline{c}} \, {\cal H}_{\underline{ab}]n}
   -2 {\cal F}_{[\underline{ab}}{}^m\, {\cal H}_{\underline{cn}]m}
  \Bigr) \beta^{nd} \;, \\[2mm]
 {\rm III}:\hspace{5pt} 0= & \Bigl( \beta^{cm} \partial_m \mathcal F_{ab}{}^d + 2 \partial_{[ \ul a}\, \mathcal Q_{\ul b]}{}^{cd}
 - \mathcal H_{mab} \mathcal R^{mcd} - \mathcal F_{ab}{}^m \mathcal Q_{m}{}^{cd} + 4 \mathcal Q_{[\ul a}{}^{[\ul cm}
 \mathcal F_{m \ul b]}{}^{\ul d]} \Bigr) \\
 &\hspace{15pt}+\Bigl( \beta^{cm}\partial_m \mathcal H_{abn} - 2 \partial_{[\ul a} \mathcal F_{\ul b]n}
 {}^{c} - 3 \mathcal H_{m[\ul a\ul b} \mathcal Q_{\ul n]}{}^{mc} + 3
 \mathcal F_{[\ul a \ul b}{}^m \mathcal F_{m\ul n]}{}^c \Bigr) \beta^{nd} \;, \\[2mm]
 {\rm IV}:\hspace{5pt} 0 = & \Bigl( -\partial_a \mathcal R^{bcd} - 2 \beta^{[\ul c m} \partial_m \mathcal Q_a{}^{\ul b] d}
 + 3 \mathcal Q_a{}^{[\ul b m}\mathcal Q_m{}^{\ul c \ul d]}
 - 3 \mathcal F_{am}{}^{[\ul b} \mathcal R^{\ul c \ul d]m} \Bigr) \\
 & \hspace{15pt}+\Bigl( 2 \beta^{[\ul c m} \partial_m \mathcal F_{an}{}^{\ul b]} - \partial_a
 \mathcal Q_{n}{}^{bc}
 + \mathcal Q_m {}^{bc} \mathcal F_{an}{}^m \\
 & \hspace{130pt} + \mathcal R^{bcm} \mathcal H_{man} 
 -4 \mathcal Q_{[\ul a}{}^{[\ul b m}\mathcal F_{m \ul n]}{}^{\ul c]} \Bigr) \beta^{nd} \;, \\[2mm]
 {\rm V}:\hspace{5pt}   0 = & \Bigl( \beta^{[\underline{c} m}\, \partial_{m}  {\cal
        R}^{\underline{ab}]d} -
     2 {\cal R}^{[\underline{ab}m}\, {\cal Q}_m{}^{\underline{cd}]} 
\Bigr)\\
& \hspace{15pt}+  \Bigl( \beta^{[\underline{c} m}\, \partial_{m}  {\cal
        Q}_n{}^{\underline{ab}]} + {\cal R}^{[\underline{ab}m}\,
        {\cal F}_{mn}{}^{\underline{c}]}+
     {\cal Q}_m{}^{[\underline{ab}} \, {\cal Q}_n{}^{\underline{c}]m}
\Bigr) \beta^{nd} \;.
\\[-6mm]
}

%%%%%%%%%%%%%%%%%%%%%%%%%%%%%%%%%%%%%%%%%%%%%%%
%%%%%%%%%%%%%%%%%%%%%%%%%%%%%%%%%%%%%%%%%%%%%%%

\pagebreak
\subsubsection*{Remarks}

We close this section with some remarks.
\begin{itemize}

\item Note that for vanishing fluxes $H$ and $f$, the equations IV and V reduce to the Bianchi
identities \eqref{bianchia} in the geometric basis. Furthermore, equation III reduces to 
$\partial_{[ \ul a}\, \mathcal Q_{\ul b]}{}^{cd}=0$. 

\item Equation II is the Bianchi identity for the usual Riemann curvature tensor,
while in the case of vanishing geometric flux equation IV is the Bianchi identity
of the second curvature tensor $\check{R}^i{}_{jkl}$ 
defined in \cite{Andriot:2012wx,Andriot:2012an}.

\item For constant fluxes, the five Bianchi identities above reduce to the
system of Bianchi identities derived in \cite{Shelton:2005cf,Ihl:2007ah,Aldazabal:2008zza} which, in our notation, read
\eq{
\label{constant fluxes}
0 &= {\cal H}_{k[\underline{ab}}\, {\cal F}^k{}_{\underline{cd}]} \;,   \\
0 &= {\cal H}_{k[\underline{ab}}\, {\cal Q}_{\underline{c}]}{}^{kj} -{\cal F}^j{}_{k[\underline{a}}\, {\cal F}^{k}{}_{\underline{bc}]} \;,   \\
0 &= {\cal H}_{kab}\, {\cal R}^{kcd} + {\cal F}^k{}_{ab}\, {\cal Q}_k{}^{cd} -
4{\cal F}^{[\underline{c}}{}_{k[\underline{a}} 
\, {\cal Q}_{\underline{b}]}{}^{\underline{d}]k} \;,    \\
0 &= {\cal F}^{[\underline{a}}{}_{ki}\, {\cal R}^{\underline{bc}]k} - {\cal Q}_i{}^{k[\underline{a}}\,{\cal Q}_k{}^{\underline{bc}]} \;,  \\
0 &= {\cal Q}_k{}^{[\underline{ab}}\, {\cal R}^{\underline{cd}]k} \; .
}

\end{itemize}

%%%%%%%%%%%%%%%%%%%%%%%%%%%%%%%%%%%%%%%%%%%%%%%
%%%%%%%%%%%%%%%%%%%%%%%%%%%%%%%%%%%%%%%%%%%%%%%
%%%%%%%%%%%%%%%%%%%%%%%%%%%%%%%%%%%%%%%%%%%%%%%
%%%%%%%%%%%%%%%%%%%%%%%%%%%%%%%%%%%%%%%%%%%%%%%
%%%%%%%%%%%%%%%%%%%%%%%%%%%%%%%%%%%%%%%%%%%%%%%
%%%%%%%%%%%%%%%%%%%%%%%%%%%%%%%%%%%%%%%%%%%%%%%
%%%%%%%%%%%%%%%%%%%%%%%%%%%%%%%%%%%%%%%%%%%%%%%
%%%%%%%%%%%%%%%%%%%%%%%%%%%%%%%%%%%%%%%%%%%%%%%

\section{Lie and Courant algebroids}
\label{sec:algebrodis}

In the previous two sections, we have considered a framework
based on the bi-vector field  $\beta$ 
to compute Bianchi identities.
However, it has been noted in the literature that the underlying mathematical structures
for non-geometric fluxes
are Lie and Courant algebroids, which generalize Lie algebras such that
structure ``constants'' become space-time dependent.

In this section, we first briefly review the relevant notions for algebroids
and then show how our previous analysis fits into this scheme.
For mathematically more rigorous and complete definitions 
we would like to refer the reader to the existing literature, in particular to
\cite{Roytenberg:01,Roytenberg:02,Gualtieri:2003dx}.

%%%%%%%%%%%%%%%%%%%%%%%%%%%%%%%%%%%%%%%%%%%%%%%
%%%%%%%%%%%%%%%%%%%%%%%%%%%%%%%%%%%%%%%%%%%%%%%

\subsubsection*{Gerstenhaber algebra and Schouten--Nijenhuis  bracket}

Let us start by introducing the \emph{Gerstenhaber algebra} which is a graded, associative,
super-commutative algebra  $G^{\star} = \bigoplus_k
G^k$ with respect to a product $\wedge$, together with a graded Lie bracket $[\cdot,\cdot]_G$ such that the
following Leibniz rule 
holds
\begin{equation}
\label{ger_leibniz}
  [ a, b\wedge c ]_G = [ a, b ]_G \wedge c + (-1)^{(k-1)l}\, b\wedge [ a, c ]_G  \;,
\end{equation}
where $a\in G^k, b \in G^l$ and $c \in G^{\star}$. 
Super-commutativity means 
\begin{equation}
\label{ger_comm}
[ a,b ]_G = - (-1)^{(k-1)(l-1)}\, [ b, a ]_G \;.
\end{equation}
In addition, for a Gerstenhaber algebra the graded Jacobi identity is satisfied
\begin{equation}
[ a, [ b, c ]_G ]_G = [[  a, b ]_G , c ]_G + (-1)^{(k-1)(l-1)}
\, [ b, [ a, c ]_G]_G\; . 
\end{equation}
An example of a Gerstenhaber algebra is the
\emph{Schouten--Nijenhuis bracket} $[\cdot ,\cdot ]_{SN}$, which 
for functions $f,g\in\mathcal C^{\infty}(M)$  and vector fields $X,Y\in \Gamma(TM)$ 
it is defined by
\eq{
  \label{sn_bracket}
[f,g]_{SN} = 0  \;, \hspace{40pt}
[X,f]_{SN} = X(f) \;,  \hspace{40pt} 
[X,Y]_{SN} = [X,Y]_L \;,
}
with $[\cdot , \cdot ]_L$ being the Lie bracket.
The Schouten--Nijenhuis bracket is uniquely  extended  to arbitrary alternating multi-vector fields 
in $\Gamma(\Lambda^\star TM)$ with usual exterior product by demanding \eqref{ger_leibniz} and 
\eqref{ger_comm}. Similarly, any exterior algebra of a Lie algebra is a Gerstenhaber algebra.

%%%%%%%%%%%%%%%%%%%%%%%%%%%%%%%%%%%%%%%%%%%%%%%
%%%%%%%%%%%%%%%%%%%%%%%%%%%%%%%%%%%%%%%%%%%%%%%

\subsubsection*{Lie algebroids}

Next, we turn to Lie algebroids.
A vector bundle $E$ over a manifold $M$ is called a \emph{Lie algebroid}, 
if it is equipped with
a Lie bracket $[\cdot,\cdot]_E$ and a bundle homomorphism $\rho : E
\rightarrow TM$, called an {\it anchor}, 
such that the following Leibniz rule  holds
\begin{equation}
\label{Leibniz}
[s_1,fs_2]_E = f[s_1,s_2]_E + (\rho(s_1)f)\, s_2 \;,
\end{equation}
where $s_i$ are sections of $E$ and $f \in C^{\infty}(M)$. 
The Lie algebroid $(E,[\cdot,\cdot]_E,\rho)$ then has the following important properties.
First, the space of sections $\Gamma(\Lambda^{\star}E)$ is a
Gerstenhaber algebra with the bracket determined by 
\begin{equation}
[ f, g ]_G = 0 \;, \hspace{40pt} 
[ f , s ]_G = -\rho(s)\,f \;, \hspace{40pt}
[s_1, s_2 ]_G = [s_1, s_2]_E \;,
\end{equation}
as well as by  \eqref{ger_leibniz}  and \eqref{ger_comm}.
Second, $\Gamma(\Lambda^{\star} E^*)$ is a 
graded differential algebra, and the differential with respect 
to the multiplication $\wedge$ is given by
\eq{
\label{algebroiddiff}
(d_E \,\omega)(s_0, \dots, s_k) =\hspace{10pt}& \sum_{i=0} ^k \;(-1)^i \rho(s_i)\left( \omega(s_0,\dots,\hat{s}_i,\dots,s_k) \right)  \\
 +&\sum_{i<j} \; (-1)^{i+j} \omega \left([s_i, s_j]_E,s_0,\dots,\hat{s}_i,\dots,\hat{s}_j,\dots,s_K \right)
 \;,
}
where $\omega \in \Gamma(\Lambda^k E^*)$ and $\{s_i\}\in \Gamma(E)$. 
Let us mention that these two properties are equivalent to the definition of a Lie algebroid given in \cite{waldmann2007poisson}.

There are two standard examples for Lie algebroids which are also of importance for our analysis. We discuss them in turn.
\begin{itemize}

\item We first consider ${\cal A}=(TM,[\cdot,\cdot]_{L},\rho= \textrm{id})$, where the anchor is
the identity map and the bracket is given by the usual Lie bracket 
    $[X,Y]_L$ of vector
  fields. The Gerstenhaber bracket on $\Gamma(\Lambda^{\star}TM)$ is
  given by the  Schouten--Nijenhuis  bra\-cket \eqref{sn_bracket} and the
  differential on $\Gamma(\Lambda^{\star}T^* M)$ is the
  usual de Rham differential. Note that  for  a vielbein basis $\{e_a\}$ 
  with geometric flux $f$ we obtain \eqref{comm_01}.

\item For the second example, we let $(M,\beta)$ be a Poisson manifold 
with Poisson tensor $\beta =  \frac{1}{2}\,\beta^{ij}e_i \wedge e_j$. 
In view of \eqref{quasipoissonb}, this means $R=\frac{1}{2}\,[\beta,\beta]_{SN}=0$. 
The Lie algebroid is then given by 
${\cal A}^*=(T^* M,[\cdot,\cdot ]_K,\rho =  \beta^\sharp)$, where 
the anchor is defined as in \eqref{anchorm}.
The bracket on $T^*M$ is the \emph{Koszul bracket} defined on one-forms as
\begin{equation}
\label{koszul}
	[\xi,\nu]_K = \mathcal{L}_{\beta^\sharp(\xi)}\nu 
 -\iota_{\beta^\sharp(\nu)}\,d\xi \; ,
\end{equation}
where the Lie derivative reads $\mathcal L_{X} = \iota_X \circ d + d \circ \iota_X$.
The associated Gerstenhaber bracket is called the \emph{Koszul--Schouten bracket}. The corresponding differential on $TM$ is given in terms of the Schouten--Nijenhuis bracket 
\begin{equation}
	d_{\beta} = [\beta,\cdot\,]_{SN} \; .
\end{equation}
Note that for the basis $\{e^a\}$ we obtain $[e^a,e^b]_K=Q_c{}^{ab}\,e^c$, with the non-geometric flux $Q$ as in \eqref{QRdef}.

\end{itemize} 
These two Lie algebroids  can be combined into a \emph{Lie
bi-algebroid} $({\cal A},{\cal A}^*)$ where, in this particular case,
the definition of the latter requires the
de Rham differential to be a derivation of the Koszul--Schouten
bracket.
That means 
\eq{
  d [ e^a, e^b]_K = [de^a,e^b]_K + [ e^a , de^b ]_K \;,
}
which can be brought into the form
\eq{
  \label{bianchi_some}
  0 = \beta^{[\ul a m} \left( \partial_{[\ul p} \, f_{\ul{qm}]}{}^{\ul b]} +
    f_{[\underline{pq}}{}^n\, f_{\underline{m}]n}{}^b \right)=
\beta^{[\ul a m} R^{\ul b]}{}_{[\ul p \ul q \ul m]} \;,
}
where $R^a{}_{bcd}$ is the curvature tensor of the connection
$\Gamma^a{}_{bc}$.
Note that due to the Bianchi identity for the curvature tensor, the relation
\eqref{bianchi_some} is automatically  satisfied.
Thus, the above two examples indeed combine into a Lie bi-algebroid.

Let us remark that the framework of Lie (bi-)algebroids provides a natural way of 
implementing \emph{non-constant} $f$- and $Q$-fluxes, which is not possible in the realm of Lie algebras.

%%%%%%%%%%%%%%%%%%%%%%%%%%%%%%%%%%%%%%%%%%%%%%%
%%%%%%%%%%%%%%%%%%%%%%%%%%%%%%%%%%%%%%%%%%%%%%%

\subsubsection*{Quasi-Lie algebroids}

In order to also describe $H$- and $R$-fluxes, we have to generalize
the above structure to include \emph{twists} (for a review  see for instance \cite{Kosmann-Schwarzbach2005}). For the two Lie algebroids of interest, twists can be realized as follows.
\begin{itemize}

\item We first consider $\mathcal{A}_H=(TM,[\cdot,\cdot]_L^H,\text{id}_{TM};H)$ with an $H$-twisted Lie bracket
\eq{
  \label{H-Lie}
	[X,Y]_L^H = [X,Y]_L - \beta^{\sharp}\left(\iota_Y\iota_X H\right)\, .
}
The bracket \eqref{H-Lie}
does not satisfy the Jacobi identity and is therefore called a 
\emph{quasi-Lie algebroid}.
However, the Jacobi identity is satisfied upon setting
$H=0$; thus $H$ measures the defect of $\mathcal A_H$ to be a Lie algebroid. 
In particular, let us evaluate \eqref{H-Lie} for two fields $e_a$
in a vielbein basis $\{e_a\}$ with geometric flux $f$. We obtain 
\eq{
   [e_a,e_b]_L^H = f_{ab}{}^p \, e_p - H_{abm}\, \beta^{mp}\, e_p = \mathcal{F}_{ab}{}^c \, e_c \, .
}

\item For the second example, we consider $\mathcal{A}_H^*=(T^*M,[\cdot ,\cdot ]_K^H,\beta^\sharp;\mathcal{R})$ with the $H$-twisted Koszul bracket
\eq{
\label{H-Koszul}
	[\xi,\nu]_K^H = [\xi,\nu]_K 
			+\iota_{\beta^\sharp(\nu)}\iota_{\beta^\sharp(\xi)} H \, .
}
Again, this bracket  does not satisfy the Jacobi identity but  
defines a quasi-Lie algebroid, and the defect of $\mathcal A^*_H$  is measured by $\mathcal R\in\Gamma(\Lambda^3 TM)$ given 
by
\eq{
	\mathcal{R}^{abc} = \tfrac{1}{2}\,[\beta,\beta]_{SN}^{abc} 
		+ \beta^{am}\beta^{bn}\beta^{ck} H_{mnk} \, .
}
That is, the Jacobi identity is satisfied iff $\mathcal R=0$, which also guarantees $\beta^\sharp$ to be an algebra homomorphism. This condition is  called the \emph{quasi-Poisson condition} \cite{Severa:2001qm}. 
Evaluating the twisted Koszul bracket \eqref{H-Koszul} for  the dual basis $\{e^a\}$, we obtain
\eq{
	[e^a,e^b]_K^H = \partial_p\beta^{ab} \, e^p + 2f_{pm}{}^{[\ul a}\beta^{m \ul b]} \, e^p 
		+\beta^{am}\beta^{bn} H_{mnp} e^p = \mathcal{Q}_c{}^{ab} \, e^c \, .
}

\end{itemize}
To summarize, we see that the fluxes $\mathcal H$
and $\mathcal R$  introduced in \eqref{redef_01} have a direct interpretation 
as the defects to the Lie algebroid properties. Furthermore, also the fluxes 
$\mathcal F$ and $\mathcal Q$ appear naturally via the brackets of two basis fields.

%%%%%%%%%%%%%%%%%%%%%%%%%%%%%%%%%%%%%%%%%%%%%%%
%%%%%%%%%%%%%%%%%%%%%%%%%%%%%%%%%%%%%%%%%%%%%%%

\subsubsection*{The associated Courant algebroid}

Our aim is to identify a framework in which \emph{all} the fluxes in \eqref{redef_01} appear. So far, we have described  $({\cal H},{\cal F})$ on $TM$ as well as $({\cal R},{\cal Q})$ on $T^*M$. 
Hence, we are  naturally lead to seek for a suitable structure on $TM\oplus T^*M$ respecting both $\mathcal{A}_H$ and $\mathcal{A}_H^*$. 
Following \cite{0885.58030,Roytenberg:01,Roytenberg:02}, we therefore consider $TM\oplus T^*M$ equipped with
\begin{itemize}
\item a bi-linear form for $(X+\xi)\in\Gamma(TM\oplus T^*M)$, where $X\in \Gamma(TM)$ and $\xi\in \Gamma(T^*M)$, which reads
\eq{
  \langle X+\xi,Y+\nu\rangle_\pm = \xi(Y)\pm \nu(X) \;,
}

\item a skew-symmetric bracket $\lb\cdot ,\cdot \rb$ on $\Gamma(TM\oplus T^*M)$ composed of
\eq{
	\lb X,Y\rb &= [X,Y]_L^H + \iota_{Y}\iota_X H  \;, \\
	\lb X,\xi\rb &= [\iota_{X},d^H]_+\,\xi - [\iota_{\xi},d_\beta^H]_+\, X 
			+ \tfrac{1}{2}\bigl(d^H-d_\beta^H\bigr)\, \langle X,\xi\rangle_- \;, \\
	\lb \xi,X\rb &= [\iota_{\xi},d_\beta^H]_+\, X - [\iota_{X},d^H]_+\,\xi  
			+ \tfrac{1}{2}\bigl(d^H-d_\beta^H\bigr)\,\langle \xi,X\rangle_- \;, \\
	\lb \xi,\nu\rb &= [\xi,\nu]_{K}^H + \iota_\nu\iota_\xi \mathcal{R} \;,
}
with $d^H$ the $H$-twisted de Rham differential given by
\eqref{algebroiddiff} via \eqref{H-Lie},  $d_{\beta}^H$ the
$H$-twisted Poisson differential
associated to \eqref{H-Koszul},  and $[\cdot,\cdot]_+$ the anti-commutator,

\item an algebra homomorphism (anchor) given by $\alpha(X+\xi)=X+\beta^\sharp(\xi)$.

\end{itemize}
This additional structure makes $TM\oplus T^*M$ into a \emph{Courant algebroid}.\footnote{Strictly speaking, this is true provided $(\mathcal{A}_H,\mathcal{A}_H^*)$ is a \emph{proto bi-algebroid} \cite{Roytenberg:01}. However, to avoid technical details  we will argue directly  that the above structure  gives a Courant algebroid. Note that an analogous construction can be made for the untwisted Lie bi-algebroid $(\mathcal{A},\mathcal{A}^*)$ leading to the untwisted version of this Courant algebroid.} 
The required axioms for the latter are the following \cite{0885.58030}:
\begin{enumerate}

\item The anchor $\alpha$ satisfies $\alpha (\lb s_1,s_2 \rb) = \lb \alpha(s_1),\alpha(s_2)\rb$ for sections $s_1,s_2\in\Gamma(TM\oplus T^*M)$.

\item The Courant bracket $\lb \cdot,\cdot\rb$ satisfies the modified Leibniz rule
\eq{
	\lb s_1,f\hspace{1pt} s_2\rb = f\hspace{1pt}\lb s_1, s_2\rb + \left(\alpha(s_1)f\right)s_2
		-\tfrac{1}{2}\langle s_1,s_2\rangle_+\, \mathcal{D}f \;,
}
where $\mathcal{D}=d^H + d_\beta^H$ and 
$f\in\mathcal C^{\infty}(M)$.

\item The anchor satisfies $\alpha\circ \mathcal D=0\,$.

\item For $s_1,s_2,s_3 \in \Gamma(TM\oplus T^*M)$ the following relation holds
\eq{
  &\alpha(s_1) \langle s_2,s_3\rangle_+ =\\
  &\hspace{40pt} \bigl\langle\lb s_1,s_2\rb + \tfrac{1}{2} \mathcal D \langle s_1,s_2\rangle_+ ,s_3\bigr\rangle_+
  + \bigl\langle s_2, \lb s_1,s_3 \rb + \tfrac{1}{2} \mathcal D \langle s_1,s_3 \rangle_+ \bigr
  \rangle_+\;.
}

\item The Jacobiator $\text{Jac}(s_1,s_2,s_3)=\lb\lb s_1,s_2\rb,s_3\rb+\text{cycl.}\,$ satisfies
\eq{
\label{jacdefects}
\text{Jac}(s_1,s_2,s_3)=\mathcal{D}\,T(s_1,s_2,s_3)
}
where $T=\frac{1}{6}\langle\lb s_1,s_2\rb,s_3\rangle_+ +\text{cycl}$.

\end{enumerate}
The first four properties are checked directly, while the last one will become apparent in the following.

To make contact with our results in section~\ref{sec:Bianchi}, we evaluate the Courant bracket $\lb\cdot,\cdot\rb$ on basis sections $\{ e_a, e^b\}\in TM\oplus T^*M$. We obtain
\eq{
	\lb e_a,e_b\rb &= \mathcal{F}_{ab}{}^c\, e_c + \mathcal{H}_{abc}\, e^c \;, \\
	\lb e_a,e^b\rb &= \mathcal{Q}_a{}^{bc} \, e_c - \mathcal{F}_{ac}{}^b \, e^c\;, \\
	\lb e^a,e^b\rb &= \mathcal{Q}_c{}^{ab} \, e^c + \mathcal{R}^{abc} \, e_c \; ,
}
which will be denoted the \emph{Roytenberg algebra} \cite{Roytenberg:01,Halmagyi:2009te}. 
Applying the anchor map to these relations
gives  the pre-Roytenberg algebra \eqref{preroytenberg}
found in the previous section.
Furthermore, evaluating the Jacobiators we find
\eq{
 \text{Jac}(e_a,e_b,e_c) =&   -3\Bigl( \partial_{[\underline{c}} \, {\cal F}_{\underline{ab}]}{}^d +
    {\cal F}_{[\underline{ab}}{}^m\, {\cal F}_{\underline{c}]m}{}^d +
    {\cal H}_{[\underline{ab}\,m}\, {\cal Q}_{\underline{c}]}{}^{md}\Bigr) e_d\\
   &-3\Bigl( \partial_{[\underline{c}} \, {\cal H}_{\underline{ab}]d}
   -2 {\cal F}_{[\underline{ab}}{}^m\, {\cal H}_{\underline{cd}]m}
  \Bigr)e^d +\tfrac{3}{2}\,\mathcal{D} {\cal H}_{abc}  \;, \\
 \text{Jac}(e_a,e_b,e^c) =& -\Bigl( \beta^{cm} \partial_m \mathcal F_{ab}{}^d 
	+ 2 \partial_{[ \ul a}\, \mathcal Q_{\ul b]}{}^{cd}
	 - \mathcal H_{mab} \mathcal R^{mcd} 
	- \mathcal F_{ab}{}^m \mathcal Q_{m}{}^{cd} 
	\\&+ 4 \mathcal Q_{[\ul a}{}^{[\ul cm}
	 \mathcal F_{m \ul b]}{}^{\ul d]} \Bigr) e_d
 	-\Bigl( \beta^{cm}\partial_m \mathcal H_{abd} 
	- 2 \partial_{[\ul a} \mathcal F_{\ul b]d}{}^{c} 
	\\&- 3 \mathcal H_{m[\ul a\ul b} \mathcal Q_{\ul d]}{}^{mc} 
	+ 3 \mathcal F_{[\ul a \ul b}{}^m \mathcal F_{m\ul d]}{}^c \Bigr)e^d
	+\tfrac{3}{2}\,\mathcal{D}\mathcal{F}^c{}_{ab}  \;, \\
\text{Jac}(e_a,e^b,e^c) =&+ \Bigl( -\partial_a \mathcal R^{bcd} 
	- 2 \beta^{[\ul c m} \partial_m \mathcal Q_a{}^{\ul b] d}
	 + 3 \mathcal Q_a{}^{[\ul b m}\mathcal Q_m{}^{\ul c \ul d]}
	\\&- 3 \mathcal F_{am}{}^{[\ul b} \mathcal R^{\ul c \ul d]m} \Bigr)e_d
	 +\Bigl( 2 \beta^{[\ul c m} \partial_m \mathcal F_{ad}{}^{\ul b]} 
	- \partial_a \mathcal Q_{d}{}^{bc}
	+ \mathcal Q_m {}^{bc} \mathcal F_{ad}{}^m 
	\\&+ \mathcal R^{bcm} \mathcal H_{mad} 
	-4 \mathcal Q_{[\ul a}{}^{[\ul b m}
	\mathcal F_{m \ul d]}{}^{\ul c]} \Bigr) e^d
	+\tfrac{3}{2}\,\mathcal{D}\mathcal{Q}_a{}^{bc}\;, \\
 \text{Jac}(e^a,e^b,e^c) =& 
	-3\Bigl( \beta^{[\underline{c} m}\, \partial_{m}  {\cal R}^{\underline{ab}]d} -
	2 {\cal R}^{[\underline{ab}m}\, {\cal Q}_m{}^{\underline{cd}]} \Bigr)e_d
	 -3  \Bigl( \beta^{[\underline{c} m}\, 
	\partial_{m}  {\cal Q}_d{}^{\underline{ab}]} \\ &
	+ {\cal R}^{[\underline{ab}m}\,
        {\cal F}_{md}{}^{\underline{c}]}+
     {\cal Q}_m{}^{[\underline{ab}} \, {\cal Q}_d{}^{\underline{c}]m}\Bigr) e^d 
     +\tfrac{3}{2}\,\mathcal{D}\mathcal{R}^{a bc}\;.
}
Let us note that 
the parenthesis  multiplying $e_d$ and $e^d$ contain 
the same terms  appearing in the four Bianchi identities 
\eqref{bianchi}. In fact, \eqref{bianchi}
can be obtained 
by applying the anchor $\alpha$ to the Jacobiators above since $\alpha\!\circ\!\mathcal{D}=0$.
Employing then \eqref{bianchi0} and \eqref{bianchi}, we can 
simplify the Jacobiators considerably and bring them into the form
\eq{
\label{jacobiators}
\text{Jac}(e_a,e_b,e_c)={\cal D}\,T(e_a,e_b,e_c)&=\tfrac{1}{2}\,{\cal D}{\cal H}_{abc} \;,\\
\text{Jac}(e_a,e_b,e^c)={\cal D}\,T(e_a,e_b,e^c)&=\tfrac{1}{2}\,{\cal D}\mathcal{F}_{ab}{}^c \; , \\
\text{Jac}(e_a,e^b,e^c)={\cal D}\,T(e_a,e^b,e^c)&=\tfrac{1}{2}\,{\cal D}\mathcal{Q}_{a}{}^{bc}\;, \\
\text{Jac}(e^a,e^b,e^c)={\cal D}\,T(e^a,e^b,e^c)&=\tfrac{1}{2}\,{\cal D}\mathcal{R}^{abc} \; .
}
These expressions are the expected defects for a Courant algebroid mentioned in \eqref{jacdefects}. We therefore have verified that the underlying structure to describe the fluxes 
$(\mathcal{H},\mathcal{F},\mathcal{Q},\mathcal{R})$ is indeed given by a Courant algebroid.

%%%%%%%%%%%%%%%%%%%%%%%%%%%%%%%%%%%%%%%%%%%%%%%
%%%%%%%%%%%%%%%%%%%%%%%%%%%%%%%%%%%%%%%%%%%%%%%
%%%%%%%%%%%%%%%%%%%%%%%%%%%%%%%%%%%%%%%%%%%%%%%
%%%%%%%%%%%%%%%%%%%%%%%%%%%%%%%%%%%%%%%%%%%%%%%
%%%%%%%%%%%%%%%%%%%%%%%%%%%%%%%%%%%%%%%%%%%%%%%
%%%%%%%%%%%%%%%%%%%%%%%%%%%%%%%%%%%%%%%%%%%%%%%
%%%%%%%%%%%%%%%%%%%%%%%%%%%%%%%%%%%%%%%%%%%%%%%
%%%%%%%%%%%%%%%%%%%%%%%%%%%%%%%%%%%%%%%%%%%%%%%

\pagebreak
\section{Conclusions}

We conclude this letter with a brief summary.
Starting from a bi-vector field $\beta$, which can be considered as the T-dual
of the Kalb-Ramond two-form $B$, we have followed a 
quite straightforward and logical path leading to 
an intricate structure for non-geometric fluxes. The approach we followed
is summarized in the diagram below.
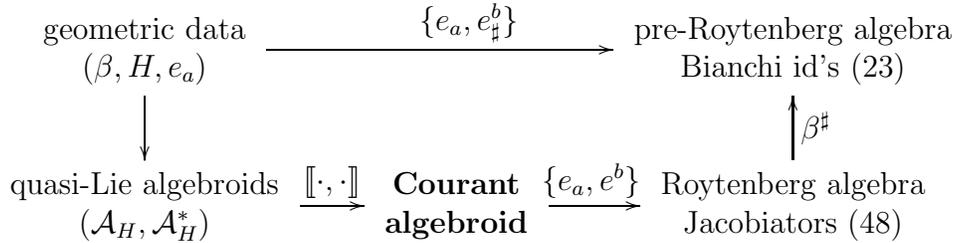
\eq{
\nonumber
\begin{xy}
\xymatrix{
{\begin{array}{c}\text{geometric data} \\ (\beta,H,e_a)\end{array}} 
\ar[rr]^{\textstyle \{e_a,e_\sharp^b\}} \ar[d]    &
& {\begin{array}{c}\text{pre-Roytenberg algebra} \\ \text{Bianchi id's \eqref{bianchi}}\end{array}}  \\
{\begin{array}{c}\text{quasi-Lie algebroids}\\(\mathcal{A}_H,\mathcal{A}_H^*)\end{array}} 
 \ar[r]^(0.6){\textstyle \lb\cdot,\cdot\rb}
&{\begin{array}{c}\text{\bf Courant} \\ \text{\bf algebroid}\end{array}} 
\ar[r]^(0.4){\textstyle \{e_a,e^b\}} 
& {\begin{array}{c}\text{Roytenberg algebra} \\ \text{Jacobiators \eqref{jacobiators}}\end{array}}
\ar[u]_{\textstyle \beta^\sharp}
}
\end{xy}
}

More concretely, we have achieved a systematic identification of the mathematical structure to
describe $(\mathcal{H},\mathcal{F},\mathcal{Q},\mathcal{R})$ fluxes in a combined way. 
It is that of a Courant
algebroid arising from twisting the standard Lie algebroid  of the
tangent and co-tangent bundle by $\cal H$. Note that the Bianchi identities \eqref{bianchi} are embedded crucially in this picture.

In view of future work, we recall again that  an effective
action for the fields $(\tilde g^{ij},\beta^{ij},\tilde\phi)$ has been derived in
\cite{Andriot:2012an}
using  double-field theory.
This action is formulated on an ordinary commutative space-time, however,
the existence of the quasi-Poisson structure and the evidence
from CFT computations suggests that there might exist an alternative
description in terms of an NCA geometry.
We hope to come back to this question in the future.

%%%%%%%%%%%%%%%%%%%%%%%%%%%%%%%%%%%%%%%%%%%%%%%
%%%%%%%%%%%%%%%%%%%%%%%%%%%%%%%%%%%%%%%%%%%%%%%
%%%%%%%%%%%%%%%%%%%%%%%%%%%%%%%%%%%%%%%%%%%%%%%
%%%%%%%%%%%%%%%%%%%%%%%%%%%%%%%%%%%%%%%%%%%%%%%
%%%%%%%%%%%%%%%%%%%%%%%%%%%%%%%%%%%%%%%%%%%%%%%
%%%%%%%%%%%%%%%%%%%%%%%%%%%%%%%%%%%%%%%%%%%%%%%
%%%%%%%%%%%%%%%%%%%%%%%%%%%%%%%%%%%%%%%%%%%%%%%
%%%%%%%%%%%%%%%%%%%%%%%%%%%%%%%%%%%%%%%%%%%%%%%

\vspace*{1.25cm}

\subsubsection*{Acknowledgements}

We thank D. L\"ust for discussion.
R.B. thanks the Simons-Center for Geometry and Physics at Stony Brook
University for hospitality.
E.P. is supported by the Netherlands Organization for Scientific Research (NWO) under the VICI grant 680-47-603.

%%%%%%%%%%%%%%%%%%%%%%%%%%%%%%%%%%%%%%%%%%%%%%%
%%%%%%%%%%%%%%%%%%%%%%%%%%%%%%%%%%%%%%%%%%%%%%%
%%%%%%%%%%%%%%%%%%%%%%%%%%%%%%%%%%%%%%%%%%%%%%%
%%%%%%%%%%%%%%%%%%%%%%%%%%%%%%%%%%%%%%%%%%%%%%%

\clearpage
%\nocite{*}
\bibliography{references}  
\bibliographystyle{utphys}

%%%%%%%%%%%%%%%%%%%%%%%%%%%%%%%%%%%%%%%%%%%%%%%
%%%%%%%%%%%%%%%%%%%%%%%%%%%%%%%%%%%%%%%%%%%%%%%
%%%%%%%%%%%%%%%%%%%%%%%%%%%%%%%%%%%%%%%%%%%%%%%
%%%%%%%%%%%%%%%%%%%%%%%%%%%%%%%%%%%%%%%%%%%%%%%

\end{document}